\documentclass{article}
\usepackage{ijcai20}
\usepackage{pdfpages}
\usepackage{times}
\usepackage{soul}
\usepackage{url}
\usepackage[hidelinks]{hyperref}
\usepackage[utf8]{inputenc}
\usepackage[small]{caption}
\usepackage{graphicx}
\usepackage{amsmath}
\usepackage{amsthm}
\usepackage{booktabs}
\usepackage{algorithm}
\usepackage{algorithmic}
\usepackage{lscape}

\newcommand{\pb}[1]{\noindent{\color{magenta} PB: #1}}

\title{A Deep Reinforcement Learning Approach to Concurrent Bilateral Negotiation}

\author{
Pallavi Bagga$^1$
\and
Nicola Paoletti$^2$\and
Bedour Alrayes$^3$\And
Kostas Stathis$^4$
\affiliations
$^{1,2,4}$Royal Holloway, University of London\\
$^3$King Saud University, Saudi Arabia\\
\emails
\{pallavi.bagga.2017$^{1}$, nicola.paoletti$^{2}$, kostas.stathis$^{4}$\}@rhul.ac.uk,
balrayes@ksu.edu.sa$^{3}$
}

\begin{document}

\maketitle

\begin{abstract}
We present a novel negotiation model that allows an agent to learn how to negotiate during concurrent bilateral negotiations 
in unknown and dynamic e-markets. The agent uses an actor-critic architecture with model-free reinforcement learning to learn a strategy expressed as a deep neural network. 
We pre-train the strategy by supervision from synthetic market data, thereby decreasing the exploration time required for learning during negotiation. As a result, we can build automated agents for concurrent negotiations that can adapt to different e-market settings without the need to be pre-programmed. Our experimental evaluation shows that our deep reinforcement learning based agents outperform two existing well-known negotiation strategies in one-to-many concurrent bilateral negotiations for a range of e-market settings. 
\end{abstract}

\section{Introduction}
We are concerned with the problem of learning a strategy for a buyer agent to engage in concurrent bilateral negotiations with unknown seller agents in open and dynamic e-markets such as E-bay\footnote{https://www.ebay.com/}. 
Previous work in concurrent bilateral negotiation has mainly focused on heuristic strategies~\cite{nguyen2004coordinating,mansour2014coordinating,an2006continuous}, some of which adapt to changes in the environment~\cite{williams2012negotiating}. Different bilateral negotiations are managed in such strategies through a coordinator agent~\cite{rahwan2002intelligent} or by coordinating multiple dialogues internally~\cite{alrayes2013agent}, but do not support agent learning which is our main focus. Other approaches use agent learning based on Genetic Algorithms (GA)~\cite{oliver1996machine,zou2014evolution}, but they require a huge number of trials before obtaining a good strategy, which makes them infeasible for online negotiation settings. Reinforcement Learning (RL)-based negotiation approaches typically employ Q-learning \cite{papangelis2015reinforcement,bakker2019rlboa,rodriguez2019context}
which does not support continuous actions. This is an important limitation in our setting because we want the agent to learn how much to concede e.g. on the price of an item for sale, which in turn naturally leads to a continuous action space. 
Consequently, the design of autonomous agents capable of learning a strategy from concurrent negotiations with other agents is still an important open problem.

We propose, to the best of our knowledge, the first Deep Reinforcement Learning (DRL) approach for one-to-many concurrent bilateral negotiations in open, dynamic and unknown e-market settings. In particular, we define a novel DRL-inspired agent model called \textit{ANEGMA}, which allows the buyer to develop an adaptive strategy to effectively use against its opponents (which use fixed-but-unknown strategies) during concurrent negotiations in an environment with incomplete information. We choose deep neural networks as they provide a rich class of strategy functions to capture the complex decisions-making behind negotiation. 

Since RL approaches need a long time to find an optimal policy from scratch 
we pre-train our deep negotiation strategies using supervised learning (SL) from a set of training examples. To overcome the lack of real-world negotiation data for the initial training, we generate synthetic datasets using the simulation environment in~\cite{alrayes2016recon} and two well-known strategies for concurrent bilateral negotiation described in ~\cite{alrayes2018concurrent} and~\cite{williams2012negotiating} respectively. 

With this work, we empirically demonstrate three important benefits of our deep learning framework for automated negotiations: 1) existing negotiation strategies can be accurately approximated using neural networks; 2) evolving a pre-trained strategy using DRL with additional negotiation experience yields strategies that even outperform the teachers, i.e., the strategies used for supervision; 3) buyer strategies trained assuming a particular seller strategy quickly adapt via DRL to different (and unknown) sellers' behaviours.


In summary, our contribution is threefold: we propose a novel agent model for one-to-many concurrent bilateral negotiations based on DRL and SL; we extend the existing simulation environment~\cite{alrayes2016recon} to generate data and perform experiments that support agent learning for negotiation; and we run an extensive experiments 
showing that our approach outperforms the existing strategies and produces adaptable agents that can transfer to a range of e-market settings.

\section{Related work}\label{relatedWork}
The existing body of automated negotiations differs from ours in one or more of the following ways: the application domain, the focus (or goal of the research), and the way and what machine learning approach has been used to improve the autonomous decision making performance of an agent.

The work in~\cite{lau2006evolutionary} uses GAs to derive a heuristic search over a set of potential solutions in order to find the mutually acceptable offers. Also, in \cite{choudhary2018evolutionary}, the authors propose a GA-based learning technique for multi-agent negotiation but with regard to making recommendations to a group of persons based on their preferences. Since we are dealing with an environment with limited information, another relevant consideration is related to RL. In \cite{bakker2019rlboa}, the authors study a modular RL based BOA (Bidding strategy, Opponent model and Acceptance condition) framework which is an extension of the work done in \cite{baarslag2016learning}. This framework implements an agent that uses tabular Q-learning to learn the bidding strategy by discretizing the continuous state/action space (not an optimal solution for large state/action spaces as it may lead to curse of dimensionality and cause the loss of relevant information about the state/action domain structure too). Q-learning is also used in~ \cite{rodriguez2019context} to provide a decision support system for the Energy market. In addition, the work in~\cite{sunder2018prosocial} uses a variable reward function for an RL approach called REINFORCE to model the pro-social or selfish behaviour of agents. Furthermore, the work of \cite{hindriks2008opponent,zeng1998bayesian} uses Bayesian Learning to learn the opponent preferences instead of the negotiation strategy. 

Previous work also consider the combination of different learning approaches to determine an optimal negotiation strategy for an agent. In \cite{zou2014evolution}, the authors propose the fusion of evolutionary algorithms (EAs) and RL that outperforms classic EAs; here the replicator dynamics is used with a GA to adjust the probabilities of strategies. In this work, the experiments have shown that different weights assigned to the historical and current payoffs (due to change in environment dynamics) while learning impact both the negotiation performance and the learning to a great extent. 
Another relevant work is \cite{lewis2017deal}, which combines SL (Recurrent Neural Network (RNN)) and RL (REINFORCE) to train on human dialogues. We also combine SL and RL but with the main focus on autonomy of negotiations rather than Natural Language Processing (NLP). Also, we differ with respect to the combination of ML approaches (i.e. Artificial Neural Network (ANN) for SL and Actor-Critic model called DDPG \cite{lillicrap2017continuous} for RL), which will be explained in subsequent sections. 

In addition and independently of the approach, numerous works in the domain of bilateral negotiation rely on the Alternating Offers protocol~\cite{rubinstein1982perfect} as the negotiation mechanism, which, despite its simplicity does not capture many realistic bargaining scenarios. 

\section{Proposed Work}\label{proposedWork}
In this section, we formulate the negotiation environment and introduce our agent negotiation model called \textit{ANEGMA} (\textit{A}daptive \textit{NEG}otiation model for e-\textit{MA}rkets). 

\subsection{Negotiation Environment}\label{negForm}
We consider e-marketplaces like E-bay where the competition is visible, i.e. a buyer can observe the number of competitors that are dealing with the same resource from the same seller. We assume that the environment consists of a single e-market $m$ with $P$ agents, with a non-empty set of buyers $B_{m}$ and a non-empty set of sellers $S_{m}$ -- these sets need not be mutually exclusive. For a buyer $b \in B_{m}$ and resource $r$, we denote with $S_{b,r}^t \subseteq S_{m}$ the set of sellers from market $m$ which, at time point $t$, negotiate with $b$ for a resource $r$ (over a range of issues $I$).  The buyer $b$ uses $|S_{b,r}^t|$ negotiation threads, in order to negotiate concurrently with each seller $\in S_{b,r}^t$. We assume that no agent can be both buyer and seller for the same resource at the same time, that is, $\forall b,r,t. \ s \in S_{b,r}^t \implies S_{s,r}^t=\emptyset$. 
$C_{b,r}^t = \{b'\neq b \in B_m \mid S_{b',r}^t \neq \emptyset \}$ is the set of competitors of $b$, i.e. those agents negotiating with the same sellers and for the same resource $r$ as that of $b$. 

As we are interested in practical settings, we adopt the negotiation protocol of~\cite{alrayes2018concurrent}, since it supports concurrent bilateral negotiations. This protocol 
assumes an open e-market environment, i.e., where agents can enter or leave the negotiation at their own will. A buyer $b$ always starts the negotiation by making an offer whose start time is $t_{\it start}$. Any negotiation is for a resource $r$, since we index the negotiation thread with the name of the seller $s$ and the resource $r$, and can last for up to time $t_b$, the maximum time $b$ can negotiate for. The deadline for $b$ is, thus, $t_{\it end} = t_{\it start} + t_b$, which for simplicity we assume for all the resources being negotiated. Information about the deadline $t_b$, Initial Price ${\it IP}_{b}$ and Reservation Price $RP_{b}$ is private to each $b \in B_m$. Each seller $s$ also has its own Initial Price ${\it IP}_{s}$, Reservation Price ${\it RP}_{s}$ and maximum negotiation duration parameter $t_s$ (which are not visible by other agents). The protocol is turn-based and allows agents to take actions from a pool $\it Actions$ at each negotiation state (from S1 to S5, see 
\cite{alrayes2018concurrent}) where
$\it Actions = \{\ offer(x),  \ \ \ reqToReserve, \ \ \  reserve, \ \ \ cancel, \ \ \ confirm, \\accept, \ \ \ exit\ \}$.

\subsection{\textit{ANEGMA} Components}
\begin{figure}
    \centering
    \includegraphics[scale=0.3]{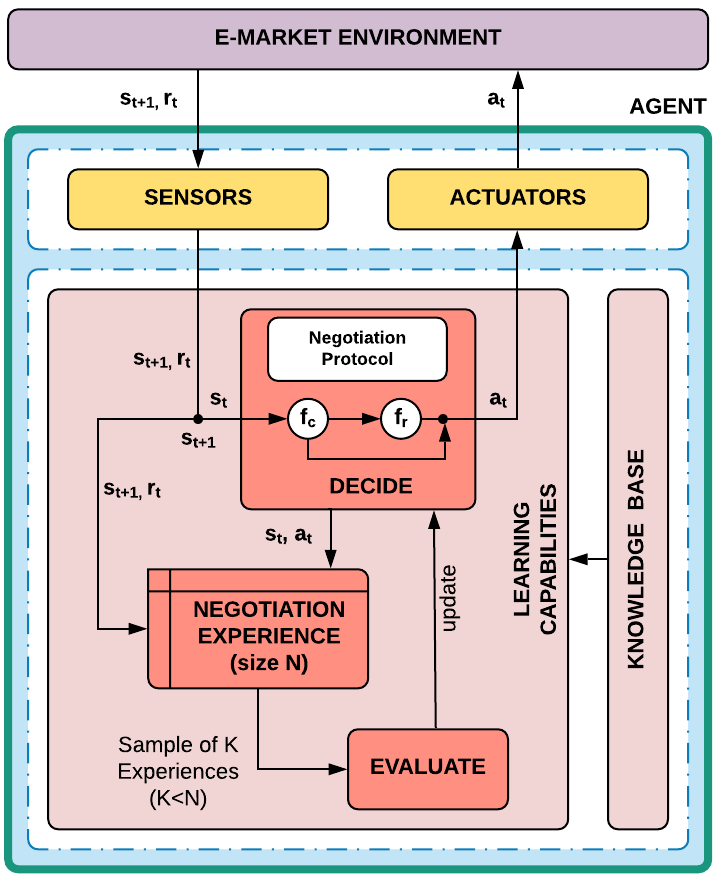}
    \caption{The Architecture of \textit{ANEGMA}}
    \label{fig:architecture}
\end{figure}

Our proposed agent negotiation model  supports learning during concurrent bilateral negotiations with unknown opponents in dynamic and complex e-marketplaces. In this model, we use a centralized approach in which the coordination is done internally to the agent via multi-threading synchronization. This approach minimizes the agent communication overhead and thus, improve the run-time performance.
The different components of the proposed model are shown in Figure \ref{fig:architecture} and explained below.
\subsubsection{Physical Capabilities:} These are the \textit{sensors} and \textit{actuators} of the agent that enable it to access an e-marketplace. More specifically, they allow a buyer $b$ to perceive the current (external) state of the environment $s_t$  and represent that state locally in the form of internal attributes as shown in Table \ref{attributes}. Some of these attributes ($NS_r$, $NC_r$) are perceived by the agent using its sensors, some of them ($IP_b$, $RP_b$, $t_{\it end}$) are stored locally in its knowledge base and some of the them ($S_{\it neg}$, $X_{\it best}$, $T_{\it left}$) are obtained while interacting with other seller agents during a negotiation. At time $t$, the internal agent representation of the environment is $s_t$, which is used by the agent to decide what action $a_t$ to execute using its \textit{actuators}. Action execution then changes the state of the environment to $s_{t+1}$.

\subsubsection{Learning Capabilities:} The foundation of our model is a component providing learning capabilities similar to those in the Actor-Critic architecture as in~\cite{lillicrap2017continuous}. 
It consists of three sub-components: \textit{Negotiation Experience}, \textit{Decide} and \textit{Evaluate}. 

\textit{Negotiation Experience} stores historical information about previous  \textit{negotiation experiences} which involve the interactions of an agent with other agents in the  market. Experience elements are of the form $\langle s_t, a_t, r_t, s_{t+1} \rangle$, where $s_t$ is the state of the e-market environment, $a_t$ is action performed by $b$ at $s_t$, $r_t$ is scalar reward or feedback received from the  environment and $s_{t+1}$ is new e-market state after executing $a_t$. 

\textit{Decide} refers to a negotiation strategy which helps $b$ to choose an optimal action $a_t$ among a set of actions ($Actions$) at a particular state $s_t$. In particular, it consists of two different functions $f_c$ and $f_r$. $f_c$ take state $s_t$ as an input and returns a discrete action among \textit{counter-offer, accept, confirm, reqToReserve} and \textit{exit}, see~\eqref{eq:classification}. When $f_c$ decides to perform a \textit{counter-offer} action, $f_r$ is used to compute, given an input state $s_t$, the value of the counter-offer, see~\eqref{eq:regression}. From a machine learning perspective, deriving $f_c$ corresponds to a classification problem, deriving $f_r$ to a regression problem.
\begin{equation}\label{eq:classification}
    f_c(s_t) = a_t, a_t \in Actions
\end{equation}
\begin{equation}\label{eq:regression}
    f_r(s_t) = x, x \in [IP_{b}, RP_{b}]
\end{equation}

\textit{Evaluate} refers to a critic which helps $b$ learn and evolve the negotiation strategy for unknown and dynamic environments. More specifically, it is a function of random $K$ ($K<N$) past negotiation experiences fetched from the database. Also, the learning process of $b$ is \textit{retrospective} since it depends on the feedback or scalar reward $r_t$ (and $r_t'$) obtained from the e-market environment by performing action $a_t$ at state $s_t$ which is calculated using \eqref{classificationReward} and \eqref{regressionReward} to evaluate the discrete and continuous action made by \textit{Decide} component at time $t$ respectively. Our design of reward functions accelerate agent learning by allowing $b$ to receive rewards after every action it performs in the environment instead of at the end of the negotiation.
\begin{equation}
\label{classificationReward}
\resizebox{.91\linewidth}{!}{$
    r_t \text{ (during classification)} = 
    \begin{cases}
    U_b(x,t), & \text{if } t \leq t_{\it end}, \text{ Agreement} \\
    -1, & \text{if } t \leq t_{\it end}, \text{ No Deal} \\
    r_t' & \text{if } a_t = \text{Counter-offer} \\
    0, & \text{otherwise}
    \end{cases}
$}\end{equation}
\begin{equation}
\label{regressionReward}
\resizebox{.91\linewidth}{!}{$
   r_t' \text{ (during regression)} = 
    \begin{cases}
    U_b(x,t), & \text{if } t \leq t_{\it end}, x \leq \forall i \in O_{t} \\
    -1, & \text{if } t \leq t_{\it end}, x > \forall i \in O_{t} \\
    0, & \text{otherwise}
    \end{cases}
$}\end{equation}
In \eqref{classificationReward} and \eqref{regressionReward}, $U_b(x,t)$ refers to the utility value of offer $x$ (generated using \eqref{eq:regression}) at time $t$ and is calculated using Initial Price ($IP_b$), Reservation Price ($RP_b$), agreement offer ($x$) and temporal discount factor ($d_t \in [0,1]$) \cite{williams2012negotiating} as defined in \eqref{utility} . The parameter $d_t$ encourages $b$ to negotiate without delay. The reward function $r_t'$ in \eqref{regressionReward} helps $b$ learn that it should not offer greater than what active sellers have already offered it. $O_{t}$ refers to a list of preferred offers of $\forall s \in S_{b,r}^t$ at time $t$. 

\begin{equation}\label{utility}
    U_b(x,t) = \left(\frac{RP_{b} - x}{RP_{b} - IP_{b}}\right).\left(\frac{t}{t_{\it end}}\right)^{d_t}
\end{equation}

In our experiments, the value of $d_t$ is set to $0.6$. Higher the $d_t$ value, higher is the penalty due to delay.

\begin{table}
    \caption{Agent's State Attributes} 
    \label{attributes}
    \begin{tabular}{p{0.07\textwidth}p{0.35\textwidth}}
    \toprule
    Attribute & Description \\
    \midrule
    $NS_{r}$ & Number of sellers that $b$ is concurrently dealing for resource $r$ at time $t$ ($|S_{b,r}^t|$).\\
    $NC_r$ & Number of buyer agents competing with $b$ for resource $r$ at time $t$ ($|C_{b,r}^t|$). \\ 
    $S_{\it neg}$ & Current state of the negotiation protocol (S1 to S5 \cite{alrayes2018concurrent})).\\
    $X_{\it best}$&  Best offer made by either $b$ or $s$ in $S_{\it neg}$.\\
    $T_{\it left}$ & Time left for $b$ to reach $t_{\it end}$ after the last action of $s$.\\
    $IP_b$ & Minimum price which $b$ can offer at the start of the negotiation. \\
    $RP_b$ & Maximum price which $b$ can offer to $s$. \\
    \hline
\end{tabular}
\end{table}
\section{Materials and Methods}\label{materialsAndMethods} 
In this section, we describe the data set collected for training the SL model (used for pre-training the \textit{ANEGMA} agent), various performance measures (used for evaluating the negotiation process) and ML models (used for the learning process).

\subsection{Data set collection}
In order to collect the data set to train \textit{ANEGMA} agent using an SL model, we have used a simulation environment~\cite{alrayes2016recon} that supports concurrent negotiations between buyers and sellers. The buyers use two different strategies presented in~\cite{alrayes2018concurrent} and~\cite{williams2012negotiating}; whereas the sellers use the strategies described in~\cite{faratin1998negotiation}. We could have also collected the negotiation examples for training using other buyer strategies for concurrent negotiation which can deal with same environment as ours, or any real-world market data; however, to the best of our knowledge none of these had readily available implementations. We have selected the input features for the dataset manually, and this set of features correspond to the agent's state attributes in Table \ref{attributes}. To avoid choosing overlapping features, we have then applied the \textit{Pearson Correlation coefficient}~\cite{lee1988thirteen} and ensured no correlation (with all correlation coefficients between $-0.16$ and $0.16$; most are closer to $0$) between the selected features. 
\subsection{Performance Evaluation Measures}
To successfully evaluate the performance of \textit{ANEGMA} and compare it with other negotiation approaches, it is necessary to identify the appropriate performance metrics. 
For our experiments, we have used the following widely adopted metrics \cite{williams2012negotiating,faratin1998negotiation,nguyen2004coordinating,alrayes2018concurrent}: \emph{Average utility rate ($U_{\it avg}$)}, 
\emph{Average negotiation time ($T_{\it avg}$)} and \emph{Percentage of successful negotiations ($S_{\%}$)}, 
which are described in Table \ref{table: performance}.

Our main motive behind calculating the $U_{\it avg}$ is to calculate the agent profit over only successful negotiations, hence we exclude the unsuccessful ones in this metric. We capture the (un)successful negotiations in a separate metric called $S_{\%}$.

\begin{table*} 
    \centering
    \caption{Performance Evaluation Metrics }
    \label{table: performance}
    \begin{tabular}{p{0.05\textwidth} p{0.8\textwidth} p{0.1\textwidth}}
        \toprule
        Metric & Definition & Ideal Value\\
        \midrule
        $U_{\it avg}$ & Sum of all the utilities of the buyer averaged over the successful negotiations. & High(1.0)\\
        
        
        $T_{\it avg}$ & Total time taken by the buyer (in milliseconds) averaged over all successful negotiations to reach the agreement. & Low($\approx$1000ms)\\
        
        $S_{\%}$ &  Proportion of total negotiations in which the buyer reaches an agreement successfully with one of the concurrent sellers. & High(100\%)\\
    \bottomrule  
    \end{tabular}
\end{table*}

\subsection{Methodology}
During our experiments, the buyer negotiates with fixed-but-unknown seller strategies in an e-market. Also, the competitor buyers use only a single fixed-but-unknown strategy which can be learnt by the buyers after some simulation runs. Hence, we consider our negotiation environment as \textit{fully-observable}. Following this, for our \textit{dynamic} (agents leave and enter the market at any time) and \textit{episodic} (the negotiation terminates at some point) environment, we use a \textit{model-free}, \textit{off-policy} RL approach which generates a \textit{deterministic policy} based on the \textit{policy gradient} method to support continuous control. More specifically, we use the \textit{Deep Deterministic Policy Gradient algorithm (DDPG)}, which is an actor-critic RL approach and generates a deterministic action selection policy for the buyer (see~\cite{lillicrap2017continuous} for more details, due to lack of space). We consider a \textit{model-free RL} approach because our buyer is more concerned with determining what action to take given a particular state rather than predicting a new state of the environment. This is because the strategies of sellers and competitor buyers are unknown in the environment. On the other hand, we consider the \textit{off-policy} approach for efficient and independent exploration of continuous action spaces. Furthermore, we, instead of initializing the RL policy randomly, use a policy generated by an Artificial Neural Network (ANN) \cite{goodfellow2016deep} due to its compatibility with DRL in order to speed up and reduce the cost of the RL process. To reduce the over-fitting and generalization errors, we also apply regularization techniques (dropout) during the training of the neural network. 
\section{Experimental Setup and Results}\label{Discussions}
We use \textit{ANEGMA} to build autonomous buyers that negotiate against unknown opponents 
in different e-market settings. Our experiments make the following hypotheses.

\textbf{Hypothesis A: }The \textit{Market Density ($\it MD$)}, the \textit{Market ratio} or \textit{Demand/Supply Ratio} ($\it MR$), the 
\textit{Zone of Agreement} ($\it ZoA$) and the \textit{Buyer's Deadline ($t_{\it end}$})
have a considerable effect on the success of negotiations. Here,
\begin{itemize}
    \item $\it MD$ is the total agents in the e-market at any given time dealing with the same resource as that of our buyer.
    \item $\it MR$ is the ratio of the total number of buyers over the sellers in the e-market. 
    \item $\it ZoA$ refers to the intersection between the price ranges of buyers and sellers for them to agree.
\end{itemize}
In practice, buyers have no control over these parameters except the deadline, which can be decided by the user 
or constrained by a higher-level goal the buyer is trying to achieve.

\textbf{Hypothesis B:} The \textit{ANEGMA} buyer outperforms SL, CONAN, and Williams' negotiation strategies in terms of $U_{\it avg}$, 
$T_{\it avg}$ and $S_{\%}$ in a range of e-market settings. 

\textbf{Hypothesis C:} An \textit{ANEGMA} buyer if trained against a specific seller strategy, still performs well against other fixed-but-unknown seller strategies. This shows that the \textit{ANEGMA} agent behaviour is \textit{adaptive} in that the agent transfers knowledge from previous experience to unknown e-market settings.

\subsection{Design of the Experiments} \label{design}
To carry out our experiments, we have extended the simulation environment RECON~\cite{alrayes2016recon} with a new online learning component for \textit{ANEGMA}.
\subsubsection{Seller Strategies}
For the purpose of training our SL model and conducting large-scale quantitative evaluations, we have used two groups of fixed seller strategies developed by Faratin \textit{et al.}~ \shortcite{faratin1998negotiation}: Time-Dependent  (\textit{Linear}, \textit{Conceder} and \textit{Boulware}) and Behaviour-Dependent (\textit{Relative tit-for-tat}, \textit{Random Absolute tit-for-tat} and \textit{Averaged tit-for-tat}). Each seller's deadline is assumed to be same as that of buyer but private to the seller. Other parameters such as ${\it IP}_s$ and ${\it RP}_s$ are determined by the $\it ZoA$ parameter, as shown in Table~\ref{table: simparam}.
\subsubsection{Simulation Parameters}
We assume that the buyer negotiates with multiple sellers concurrently to buy a second-hand laptop ($r = Laptop$) based only on a single issue \textit{Price} ($I = \{Price\}$). We stress that the single-issue assumption is realistic in several real-world e-markets. 
The simulated market allows the agents to enter and leave the market at their own will. The maximum number of agents allowed in the market, the demand/supply ratio, the buyer's deadline and the $\it ZoA$s are simulation-dependent.

As in \cite{alrayes2018concurrent}, three qualitative values are considered for each parameter during simulations, e.g., High (H), Average (A) and Low (L) for $\it MD$ or Long (Lg), Average (A) and Short (Sh) for $t_{\it {end}}$. Parameters are reported in Table~\ref{table: simparam}. The user can select one of such qualitative values for each parameter. Each qualitative value corresponds to a set of three quantitative values, of which only one is chosen at random for each each simulation (e.g., setting $H$ for parameter $\it MD$ corresponds to choosing at random among $30$, $40$, and $50$). The only exception is parameter $\it ZoA$, which maps to a range of uniformly distributed quantitative values for the seller's initial price ${\it IP}_s$ and reservation price ${\it RP}_s$ (e.g., selecting $A$ for $\it ZoA$ leads to a value of ${\it IP}_s$ uniformly sampled in the interval $[580,630]$). 
Therefore, the total number of simulation settings is 81, as we consider $3$ possible settings for each of $\it MD$, $\it MR$, $\it t_{end}$, and $\it ZoA$ (see Table~\ref{table: simparam}). 
\begin{table}
  \caption{Simulation Parameter Values} 
  \label{table: simparam}
  \begin{tabular}{p{0.03\textwidth}p{0.42\textwidth}}
    \toprule
     & Values  \\
    \midrule
    $IP_b$ & $[300-350]$ \\ 
    $RP_b$ & $[500-550]$  \\
    $IP_s$ & $100\% [500-550], 60\% [580-630], 10\% [680-730]$\\
    $RP_s$ & $100\% [300-350], 60\% [380-430], 10\% [480-530]$\\
    $\it MD$ & H$\{30, 40, 50\}$, A$\{18, 23, 28\}$, L$\{8, 10, 12\}$\\
    $\it MR$ & H$\{10$:$1,1$:$1,1$:$10\}$, A$\{5$:$1,1$:$1,1$:$5\}$, L$\{2$:$1,1$:$1,1$:$2\}$\\
    $t_{\it end}$ & Lg$[151$s \textendash $210$s$]$, A$[91$s \textendash $150$s$]$, Sh$[30$s \textendash $90$s]\\ 
    $ZoA$ & H($100$\%), A($60$\%), L($10$\%)\\
    \bottomrule
  \end{tabular}
\end{table}
\subsection{Empirical Evaluation}
We evaluate hypotheses A, B and C as described at the beginning of this section.

\subsubsection{\textbf{Hypothesis A} \textit{($\it MD$, $\it MR$, $\it ZoA$ and $t_{\it end}$ have significant impact on negotiations)}}
\begin{figure}
    \centering
    \includegraphics[scale = 0.43]{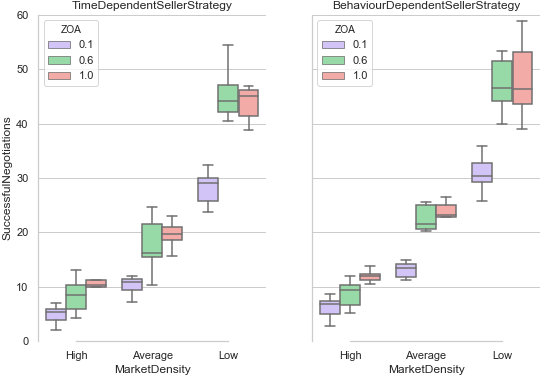}
    \caption{Effect of Market Density ($\it MD$) and Zone of Agreement ($\it ZoA$) on Proportion of Successful Negotiations ($\it S_{\%}$) using time-dependent strategies (left) and behaviour-dependent strategies (right).}
    \label{fig:hypA}
\end{figure}

\begin{figure}
    \centering
    \includegraphics[scale = 0.43]{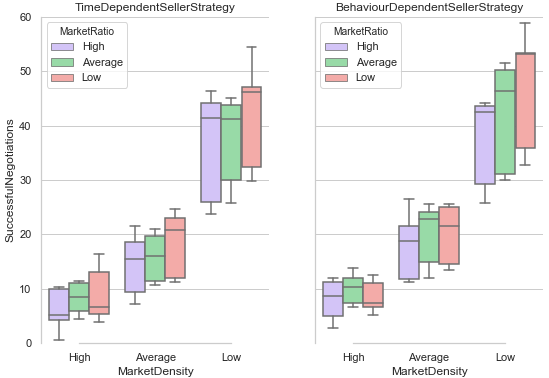}
    \caption{Effect of Market Density ($\it MD$) and Market Ratio ($\it MR$) Proportion of Successful Negotiations ($\it S_{\%}$) using time-dependent strategies (left) and behaviour-dependent strategies (right).}
    \label{fig:hypA1}
\end{figure}

We experimented with $81$ different e-market settings, by considering, for each setting, both time-dependent and behaviour-dependent seller strategies over $500$ simulations using the CONAN buyer strategy. As shown in Figure \ref{fig:hypA}, these experiments suggest that $\it MD$ and $\it ZoA$ have a considerable effect on $S_{\%}$. From our observations, when $\it MD$ is low, the agents reach more negotiation agreements. Also, there is not much difference in the agreement rate for $60$\% $\it ZoA$ and $100$\% $\it ZoA$ when $\it MD$ is low. The very low number of successful negotiations for $10$\% $\it ZoA$ is not unexpected since only a minority of agents is willing to concede more in such a small $\it ZoA$. On the other hand, $\it MR$ and $t_{\it end}$ have, according to our experiments, a comparably minor impact on the negotiation success (only some effect of $\it MR$ on $S_{\%}$ is observed under behaviour-dependent strategies and low $\it MD$ as shown in Figure \ref{fig:hypA1}).
These results support our hypothesis.

\subsubsection{\textbf{Hypothesis B} \textit{(ANEGMA outperforms SL and CONAN)}} We performed simulations for our \textit{ANEGMA} agent in low $\it MD$, 60\% and 100\% $ZoA$, high $\it MR$ and a long $t_{\it end}$ because these settings yielded the best performance in terms of $S_{\%}$ in our experiments for Hypothesis A. We have used these settings against \textit{Conceder Time Dependent} and \textit{Relative Tit for Tat Behaviour Dependent} seller strategies. 
Firstly, we collected training data for our SL approach (ANN) using two distinct strategies for supervision, viz. CONAN~\cite{alrayes2018concurrent} and Williams~\cite{williams2012negotiating}. Both were run for $500$ simulations and with the same settings. Table~\ref{table: conan} compares the performances of CONAN's and Williams' models.  CONAN outperforms Williams' strategy in these settings. 


\begin{figure}
    \centering
    \includegraphics[scale = 0.43]{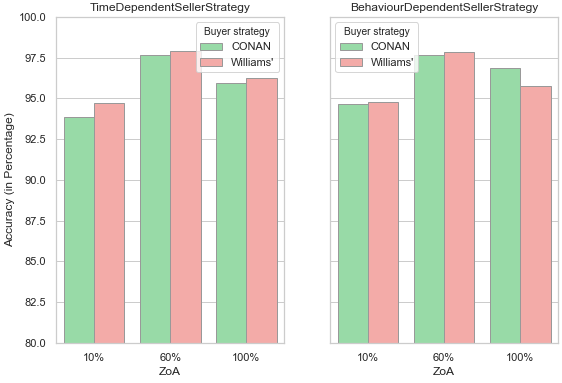}
    \caption{Training Accuracy's of ANN when trained using datasets collected by negotiating CONAN and Williams' buyer strategy (for different $\it ZoA$s) against time-dependent strategies (left) and behaviour-dependent strategies (right).}
    \label{fig:accuracies}
\end{figure}

\begin{table*}
  \caption{Performance comparison of CONAN and Williams' model. Best results are in bold.}  
  \label{table: conan}
  \begin{tabular}{p{0.04\textwidth} p{0.19\textwidth}p{0.22\textwidth} p{0.19\textwidth}p{0.20\textwidth}}
    \toprule
    Metric & \multicolumn{2}{l}{CONAN}  & \multicolumn{2}{l}{Williams'}   \\
    \midrule
     \multicolumn{5}{c}{\textbf{\textit{Conceder Time Dependent Seller Strategy}}} \\
   
    & 60\% $\it ZoA$ & 100\%  $\it ZoA$  & 60\% $\it ZoA$ & 100\%  $\it ZoA$ \\
   
    $U_{\it avg}$  &
    \textbf{0.27 $\pm$ 0.03} & 
     0.25 $\pm$ 0.07 &
    0.18 $\pm$ 0.08  &
     0.17 $\pm$ 0.04 \\

    $T_{\it avg}$ &
    \textbf{172942.78 $\pm$ 15177.77} &
    174611.43 $\pm$ 15139.52 &
    177091.09  $\pm$ 15304.90 &
     174468.311 $\pm$ 15365.11\\
    
    $S_{\%}$ &
    \textbf{80.76 }&  
     79.08 & 
    78.21 &
    78.05
    \\
    
    \multicolumn{5}{c}{\textbf{\textit{Relative Tit For Tat Behaviour Seller Strategy}}} \\
    
    $U_{\it avg}$  &
    \textbf{0.25 $\pm$ 0.03 }&
   0.24 $\pm$ 0.04 &
   0.22 $\pm$ 0.05 &
   0.21 $\pm$ 0.06\\
    
    $T_{\it avg}$ &
    \textbf{175198.93 $\pm$ 14193.23} &
    179529.47 $\pm$ 14651.15 & 
    176334.65 $\pm$ 14683.03 &
    176468.31 $\pm$ 15365.11 \\
    
    $S_{\%}$ &
    \textbf{80.69 } 
    & 79.90
    & 73.00 &
     73.21 \\
    \bottomrule
  \end{tabular}
\end{table*}

\begin{table*}
    \caption{Performance comparison of ANN VS ANEGMA(SL+RL) VS ANEGMA(RL) when $\it ZoA$ is 60\%. Best results are in bold. ANN-C and ANN-W correspond to ANN trained using data set collected from CONAN and Williams' approach respectively, whereas ANEGMA(SL+RL)-C and ANEGMA(SL+RL)-W correspond to ANEGMA(DDPG) initialized with ANN-C and ANN-W respectively.} 
    \label{annvsanegma1}
    \begin{tabular}{p{0.04\textwidth} p{0.182\textwidth} p{0.182\textwidth}p{0.182\textwidth}p{0.186\textwidth} p{0.14\textwidth}}
    \toprule
    Metric & \multicolumn{2}{l}{ANN} & \multicolumn{2}{l}{ANEGMA(SL+RL)} & {ANEGMA(RL)} \\
    \midrule
    \multicolumn{6}{c}{\textbf{\textit{Trained and Tested on Conceder Time Dependent Seller Strategy}}} \\
    
    & \textit{ANN-C} & \textit{ANN-W} & \textit{ANEGMA(SL+RL)-C} & \textit{ANEGMA(SL+RL)-W} &  \\
    $U_{\it avg}$  & 
    0.27 $\pm$ 0.04 & 0.21 $\pm$ 0.08 &
    \textbf{0.29 $\pm$ 0.04} & 0.21 $\pm$  0.04 &
    -0.38 $\pm$ 0.14 \\
    
    $T_{\it avg}$  &
    173529.47 $\pm$ 14651.15 & 171096.09 $\pm$ 14584.90&
    67750.62$\pm$ 37628.57 & 132477.71 $\pm$ 26601.48 &
    \textbf{768.55} $\pm$ \textbf{373.65} \\
    
   $S_{\%}$&
   80.80 & 80.34 &
   \textbf{87.12} & 81.72 &
   64.54 \\  
   
   \multicolumn{6}{c}{\textbf{\textit{Trained and Tested on Relative Tit for Tat Behaviour Dependent Seller Strategy}}} \\ 
   
    & \textit{ANN-C} & \textit{ANN-W} & \textit{ANEGMA(SL+RL)-C} & \textit{ANEGMA(SL+RL)-W} &  \\

      $U_{\it avg}$  &
    0.26 $\pm$ 0.03 & 0.23 $\pm$ 0.05 &
    \textbf{0.29 $\pm$  0.03} & 0.23 $\pm$ 0.14 &
    -0.19 $\pm$ 0.42 \\

      $T_{\it avg}$ &
    176018.69 $\pm$ 14380.28 & 169334.65 $\pm$ 12389.03 &
    36331.34 $\pm$  70247.33  & 41225.17 $\pm$ 72938.79 &
    \textbf{755.74} $\pm$ \textbf{292.29}\\
    
      $S_{\%}$ &
    81.86 & 74.80 &
    \textbf{86.03} & 74.57 &
    61.51 \\
    
    \hline
\end{tabular}
\end{table*}

\begin{table*}
    \caption{Performance comparison of ANN VS ANEGMA(SL+RL) VS ANEGMA(RL) when $\it ZoA$ is 100\%. Best results are in bold. ANN-C and ANN-W correspond to ANN trained using data set collected from CONAN and Williams' approach respectively, whereas ANEGMA(SL+RL)-C and ANEGMA(SL+RL)-W correspond to ANEGMA(DDPG) initialized with ANN-C and ANN-W respectively.} 
    \label{annvsanegma2}
    \begin{tabular}{p{0.04\textwidth} p{0.182\textwidth} p{0.182\textwidth}p{0.182\textwidth}p{0.186\textwidth} p{0.145\textwidth}}
    \toprule
    Metric & \multicolumn{2}{l}{ANN} & \multicolumn{2}{l}{ANEGMA(SL+RL)} & {ANEGMA(RL)} \\
    \midrule
    \multicolumn{6}{c}{\textbf{\textit{Trained and Tested on Conceder Time Dependent Seller Strategy}}} \\
    
    & \textit{ANN-C} & \textit{ANN-W} & \textit{ANEGMA(SL+RL)-C} & \textit{ANEGMA(SL+RL)-W} &  \\
    $U_{\it avg}$  & 
  0.23 $\pm$ 0.04 & 
  0.17 $\pm$ 0.08 &
    \textbf{ 0.27 $\pm$ 0.51} & 
     0.21 $\pm$ 0.71 &
   -0.88 $\pm$ 0.16 \\
    
    $T_{\it avg}$  &
   172234.73 $\pm$ 14516.15 &
   170969.09 $\pm$ 14464.09&
    171266.64  $\pm$  11573.38 & 
    185425.74  $\pm$  19909.06 &
    \textbf{1021.95 $\pm$ 771.47} \\
    
   $S_{\%}$&
   79.80 &
  78.49 &
   \textbf{ 79.73} & 
    74.61 &
    59.41 \\  
   
   \multicolumn{6}{c}{\textbf{\textit{Trained and Tested on Relative Tit for Tat Behaviour Dependent Seller Strategy}}} \\ 
   
    & \textit{ANN-C} & \textit{ANN-W} & \textit{ANEGMA(SL+RL)-C} & \textit{ANEGMA(SL+RL)-W} &  \\

      $U_{\it avg}$  &
    0.26 $\pm$ 0.30 &
    0.18 $\pm$ 0.55 &
    \textbf{0.29  $\pm$ 0.35 }& 
    0.23  $\pm$  0.84 &
    -0.24  $\pm$  0.55\\

      $T_{\it avg}$ &
    160178.98 $\pm$ 14809.18 &
    163943.05 $\pm$ 12895.03 &
      33695.16 $\pm$ 64292.37 & 
   23528.25 $\pm$ 61440.37 &
   \textbf{817.67$\pm$523.67} \\
    
      $S_{\%}$ &
    75.61 & 
    74.02 &
   \textbf{80.81}  & 
   72.53 &
   58.09 \\
    
    \hline
\end{tabular}
\end{table*}

\begin{table*}
    \caption{Performance comparison for the adaptive behaviour of ANN VS ANEGMA(SL+RL) VS ANEGMA(RL). Best results are in bold. ANN-C and ANN-W correspond to ANN trained using data set collected from CONAN and Williams' approach respectively, whereas ANEGMA(SL+RL)-C and ANEGMA(SL+RL)-W correspond to ANEGMA(DDPG) initialized with ANN-C and ANN-W respectively.}
    \label{conanvsannvsanegmaAdaptive}
    \begin{tabular}{p{0.04\textwidth} p{0.182\textwidth} p{0.182\textwidth}p{0.182\textwidth}p{0.182\textwidth} p{0.14\textwidth}}
    \toprule
    Metric & \multicolumn{2}{l}{ANN} & \multicolumn{2}{l}{ANEGMA(SL+RL)} & {ANEGMA(RL)} \\
    \midrule
    \multicolumn{6}{c}{\textbf{\textit{Trained on Relative Tit for Tat Behaviour Dependent and Tested on Conceder Time Dependent Seller Strategy}}} \\
    
    & \textit{ANN-C} & \textit{ANN-W} & \textit{ANEGMA(SL+RL)-C} & \textit{ANEGMA(SL+RL)-W} &  \\
   
    $U_{\it avg}$  & 
    0.16 $\pm$ 0.05 & 0.17 $\pm$ 0.04 &
    \textbf{0.26 $\pm$ 0.06} & 0.23 $\pm$  0.07 &
    -0.36 $\pm$ 0.12 \\

    $T_{\it avg}$  &
    174139.30 $\pm$ 14655.42 & 174035.91 $\pm$ 14627.59 &
    38402.78$\pm$ 64367.45 & 108051.11 $\pm$ 57755.84&
   \textbf{ 738.55} $\pm$ \textbf{279.65} \\
    
   $S_{\%}$&
   70.51 & 69.54 &
   \textbf{86.72} & 81.32 &
   54.54 \\  
   
   \multicolumn{6}{c}{\textbf{\textit{Trained on Conceder Time Dependent and Tested on Relative Tit for Tat Behaviour Dependent Seller Strategy}}} \\ 
   
    & \textit{ANN-C} & \textit{ANN-W} & \textit{ANEGMA(SL+RL)-C} & \textit{ANEGMA(SL+RL)-W} &  \\

      $U_{\it avg}$  &
    0.25 $\pm$ 0.05 & 0.21 $\pm$ 0.04 &
    \textbf{0.28 $\pm$  0.01} & 0.21 $\pm$ 0.08 &
    -0.28 $\pm$ 0.51 \\
   
      $T_{\it avg}$ &
    176048.05 $\pm$ 14423.36& 175170.19 $\pm$ 14623.53 &
    19295.84 $\pm$ 53767.54  &114510.0 $\pm$ 64667.79 &
   \textbf{806.83} $\pm$ \textbf{375.51}\\
    
      $S_{\%}$ &
    79.67 & 76.50 &
    \textbf{84.72} & 71.37 &
    51.89 \\
    
    \hline
\end{tabular}
\end{table*}

Then, the resulting trained ANN models -- called ANN-C and ANN-W respectively -- were used as the initial strategies in our DRL approach (based on DDPG), where strategies are evolved using negotiation experience from additional $500$ simulations. In the remainder, we will abbreviate this model by \textit{ANEGMA(SL+RL)}.

Finally, we use test data from $100$ simulations to compare the performance of such derived ANEGMA(SL+RL) buyers against CONAN, Williams' model, ANN-C, ANN-W, and the so-called ANEGMA(RL) model, which uses DDPG but initialized with a random strategy.


According to our results shown in Tables \ref{annvsanegma1} and \ref{annvsanegma2}, 
the performance of ANN-C is comparable to that of CONAN for both 60\% and 100\% $\it ZoA$s (see Table~\ref{table: conan}), and we observe the same for ANN-W and the Williams' strategy. 
So, we conclude that our approach can successfully produce neural network strategies which are able to imitate the behaviour and the performance of CONAN and Williams' models (moreover, the training accuracy's were in the range between $93.0\%$ and $98.0\%$ as shown in Figure \ref{fig:accuracies}). 

Even more importantly, the results demonstrate that ANEGMA(SL+RL)-C (i.e. DDPG initialized with ANN-C) and ANEGMA(SL+RL)-W (i.e. DDPG initialized with ANN-W) improve on their respective initial ANN strategies obtained by SL, and outperform the DRL agent ANEGMA(RL) initialized at random for both 60\% and 100\% $\it ZoA$s, see Tables~\ref{annvsanegma1} and~\ref{annvsanegma2}. This proves that both the evolution of the strategies via DRL and the initial supervision are beneficial. Furthermore, ANEGMA(SL+RL)-C and ANEGMA(SL+RL)-W also outperform the existing ``teacher strategies'' (CONAN and Williams) 
used for the initial supervision and hence can improve on them, see Table \ref{table: conan}. 

\subsubsection{\textbf{Hypothesis C}\textit{ (ANEGMA is adaptable)}} 
In this final test, we evaluate how well our \textit{ANEGMA} agents can adapt to environments different from those used at training-time. Specifically, we deploy strategies trained using \textit{Conceder Time Dependent} opponents into an environment with \textit{Relative Tit for Tat Behaviour Dependent} opponents, and viceversa. The ANEGMA agents use experience from 500 simulations to adapt to the new environment. 
Results are presented in Table~\ref{conanvsannvsanegmaAdaptive} for 60\% $\it ZoA$ and show clear superiority of the ANEGMA agents over the ANN-C and ANN-W strategies which, without online retraining, cannot maintain their performance in the new environment. This confirms our hypothesis that ANEGMA agents can learn to adapt at run-time to different unknown seller strategies. 



\subsubsection{\textbf{Further discussion}}
Pondering over the negative average utility values of ANEGMA(RL) (see Tables~\ref{annvsanegma1} and~\ref{annvsanegma2}), recall that we define the utility value as per Equation~\eqref{utility} but without the discount factor term. Therefore, if an agent concedes a lot to make a deal, it will collect a negative utility. This is precisely what happens to the initial random (and inefficient) strategy used in the ANEGMA(RL) configuration. The combination of SL and DRL prevents this very problem as it uses an initial pre-trained strategy which is much less likely to incur negative utility values. 

For the same reason, we observe a consistently shorter average negotiation time for ANEGMA(RL), which is caused by the buyer that concedes more to reach the agreement without negotiating for a long time with the seller. Hence, a shorter $T_{\it avg}$ alone does not generally imply a better negotiation performance.


An additional advantage of our approach is that it alleviates the common limitation of RL that an RL agent needs a non-trivial amount of experience before reaching a satisfactory performance.

\section{Conclusions and Future Work}\label{conclusionsAndFutureScope}
We have proposed \textit{ANEGMA}, a novel agent negotiation model that supports agent learning and adaptation during concurrent bilateral negotiations for a class of e-markets such as E-bay. Our approach derives an initial neural network strategy via supervision from well-known existing negotiation models, and evolves the strategy via DRL.  We have empirically evaluated the performance of {\em ANEGMA} against fixed-but-unknown seller strategies in different e-market settings, showing that {\em ANEGMA} outperforms the well-known existing ``teacher strategies'', the strategies trained with SL only and those trained with DRL only. Crucially, our model also exhibit adaptive behaviour, as it can transfer to environments with unknown sellers' behaviours different from training. 

As future work, we plan to consider more complex market settings including multi-issue negotiations and dynamic opponent strategies. 

\bibliographystyle{named}
\bibliography{output}

\end{document}